\documentclass[aps,preprint]{revtex4}%
\usepackage{amsfonts}
\usepackage{amsmath}
\usepackage{amssymb}
\usepackage{subfigure}
\usepackage{graphicx}%
\usepackage{graphicx}
\usepackage{float}

\begin{document}
	\title{Thermodynamic Instabilities of Conformal Gravity Holography in Four Dimensions}
	\author{Yue Song$^{a,b}$}
	\email{syphysics@std.uestc.edu.cn}
	\author{Benrong Mu$^{a,b,c}$}
	\email{benrongmu@cdutcm.edu.cn}
	\affiliation{$^{a}$ Center for Joint Quantum Studies, College of Medical Technology,
		Chengdu University of Traditional Chinese Medicine, Chengdu, 611137, PR China}
	\affiliation{$^{b}$ School of Physics, University of Electronic Science and Technology of China, Chengdu, 611731, China}	
	\affiliation{$^{c}$ Center for Theoretical Physics, College of Physics, Sichuan University,
Chengdu, 610064, PR China}
	\begin{abstract}
		Recently, a conjecture has been proposed, which indicates a correlation between super-entropy black holes and the thermodynamic instability \cite{Cong:2019bud}. W.Cong et al. suggested that the $C_{V}$ (specific heat capacity at constant volume) and $C_{P}$ (specific heat capacity at constant pressure) of a super-entropy black hole could not be greater than zero simultaneously as set in the extended phase space, which implies that the black hole is unstable in extended thermodynamics. This conjecture is intriguing and meaningful. Therefore, we did a study on that. After deriving the equations of specific heat capacities as well as plotting the relevant curves of the four-dimensional conformal black holes (a kind of super-entropy black hole), we obtained regions on the graphs where $C_{V}$ and $C_{P}$ are simultaneously greater than zero, which contradicts the conjecture that super-entropy black hole corresponds to its thermodynamic instability. Thus far, we have provided a counterexample to the hypothesis in \cite{Cong:2019bud}.
		
	\end{abstract}
	\keywords{}\maketitle
	\tableofcontents

	\section{Introduction}
	
	\label{Sec:Intro}
	Hitherto, the thermodynamics of black holes has drawn extensive attention especially for those who devote themselves to investigate the secrets within space-time. It is considered as a combination between quantum mechanics and the theory of relativity. Via the analogy with the thermodynamics of general system, black hole have become a sort of special thermodynamic system since the similarity. As a result, the thermodynamics of black holes has been widely applied \cite{Johnson:2019mdp,Bekenstein:1973ur,Hawking:1982dh,Chamblin:1999hg,Johnson:2014yja,Frassino:2014pha,Xu:2020gud,Mu:2015qta,Liang:2020uul,Mu:2019bim,Jacobson:1995ab,Kubiznak:2012wp,Cai:2001dz,Caldarelli:1999xj,Jacobson:1993vj,Gunasekaran:2012dq,Hayward:1997jp,Dolan:2011xt,Dolan:2010ha,Barvinsky:2002qu,Khan:2021tzv,Abbasvandi:2016oyw,Hammad:2018ddv,Faizal:2014tea,Pourhassan:2017kmm,Hendi:2016yof,Amelino-Camelia:2005zpp,Kim:2012cma,Tawfik:2015kga,Pourhassan:2016zzc,Mbarek:2018bau,Hull:2021bry,Rafiee:2021hyj,Cong:2021jgb,Ahmed:2023snm}. In this field, black holes are recognized as obeying the first law of thermodynamics i.e. $dU=TdS$. The current thermodynamics of black holes has defined the corresponding thermodynamic state parameters, such as temperature, pressure in the extended phase space, volume and so on. Particularly, the entropy of a black hole is defined as $S=\frac{A}{4}$, where $A$ is denoted as the area of horizon. Meanwhile, the mass is linked to its energy i.e. $U=M$. Using a noteworthy instance, the four-dimensional Schwarzschild black hole \cite{Argurio:1998xm,Bueno:2017qce}, we can learn $S=4\pi M^{2}$ as well as $T=\frac{1}{8\pi M}$. These indicate that with black hole generating Hawking radiation constantly, the decrease in mass-energy certainly rises the temperature, which results in more loss in mass-energy. This similar instability mentioned above can be reflected by observing a negative specific heat capacity, which is conveyed via \cite{Johnson:2019mdp}
	\begin{equation}\label{eqn:Q16}
		C=T\frac{\partial S}{\partial T}=-\frac{1}{8\pi T^{2}}. \\
	\end{equation}
	
	Within the extended phase space, the thermodynamic pressure is generally associated with the cosmological constant using $P=-\frac{\varLambda}{8\pi}$ to express, while the mass of a black hole is found to be consistent with enthalpy which satisfies $H=U+PV$ \cite{Kastor:2009wy}. As for the thermodynamic volume, it is defined as the the conjugate quantity of pressure, which is calculated by $V=(\frac{\partial M}{\partial P})_{S}$. The first law of thermodynamics could be rewritten as $dH=TdS+VdP$. For the space-time with a negative cosmological constant, it is clear that one can obtain its pressure to be positive for satisfying the construction of thermodynamics.
	
	In the ordinary investigation related to the thermodynamic stability, specific heat capacity is always considered as the specific heat capacity at constant pressure $C_{P}$ (the specific heat capacity obtained by remaining the cosmological constant unchanged). Nevertheless, for thermodynamics in the extended phase space, the specific heat capacity at constant volume $C_{V}$ also contains additional information, which should not be ignored.
	
	In order to comprehend the super-entropy, reviewing the inverse isoperimetric inequality is essential. The inverse isoperimetric inequality refers to the relationship between the area (entropy of a black hole) and the volume of a black hole, which is represented in Euclidean space $\mathbb{E}^{d-1}$ as the following ratio
	\begin{equation}\label{eqn:Q17}
		\mathcal{R}=(\frac{(d-1)V}{\omega_{d-2}})^{\frac{1}{d-1}}(\frac{\omega_{d-2}}{A}), \\
	\end{equation}
	where $\omega_{d}$ (volume of a unit d-dimensional sphere) satisfies $\omega_{d}=\frac{2\pi^{\frac{d+1}{2}}}{\Gamma(\frac{d+1}{2})}$.
	
	An inverse isoperimetric inequality $\mathcal{R}\geq1$ was postulated in \cite{Cvetic:2010jb}. At the time, almost all known black holes were discovered to obey this inequality. Thus, there were more and more scholars involving in the investigation for verifying the inverse isoperimetric inequality of various black holes. After sifting through time, plenty of black holes have emerged the violation in the inverse isoperimetric inequality i.e. $\mathcal{R}\text{<}1$. The inverse isoperimetric inequality implies that the black hole reaches the maximum in entropy at the saturation of the equation ($\mathcal{R}\text{=}1$) when it is situated at the corresponding thermodynamic volume. Hence, black holes violating the inverse inequality are assumed to possess a larger entropy than the ratio admitting, which are so-called super-entropy black hole.
	
	Despite the fact that super-entropy black holes have already been discovered, one still cannot illustrate the physical implications behind them. Recently, a related thought is to connect the violation of the inverse isoperimetric inequality with thermodynamic instability \cite{Johnson:2019mdp,Cong:2019bud}. In \cite{Johnson:2019mdp}, Clifford V. Johnson chose to investigate the charged BTZ black hole (a kind of super-entropy black hole). Meanwhile, in the process of deriving $C_{V}$, he found it was constantly negative. Therefore, Clifford V. Johnson associated the super-entropy with $C_{V}$ being negative, which led to a further extension that the violation of the inverse isometric inequality indicates a new thermodynamic instability i.e. $C_{V}<0$.
	
	Since then in \cite{Cong:2019bud}, Wan Cong et al. found there is a non-negative region of $C_{V}$ under certain conditions while the corresponding $C_{P}$ obtained by the same radius is in a negative state with detailed calculations in generalized exotic BTZ black holes. Therewith they argue that the conjecture in \cite{Johnson:2019mdp} can be further extended to reveal the relation between the violation of the inverse isometric inequality and the thermodynamic instability. Hence the manners of judging thermodynamic instability are no longer restricted to confirm $C_{V}$ is negative but to confirm that $C_{V}$ and $C_{P}$ are impossible to be simultaneously greater than zero.
	
	However, we are skeptical of the universality for the conjecture. With the  calculations of specific heat capacities in a four-dimensional conformal black hole, we find the curves do not fit well with the conjecture mentioned.
	
	As described above, we have chosen the four-dimensional conformal black hole as the specific object of investigation for verifying if there is a link between super-entropy and the thermodynamic instabilities. The four-dimensional conformal black hole, one of the super-entropy black holes, inherently violates the inverse isoperimetric inequality. Its thermodynamic properties are well defined and the explicit analytical solution of $C_{P}$ as well as $C_{V}$ can be calculated more conveniently. After obtaining the corresponding curves of specific heat capacity, we have a deeper and clearer understanding of the connection between super-entropy and thermodynamic instability.
	
	The rest of paper is structured as follows. In section \ref{Sec:CGH}, we review the thermodynamic properties of four-dimensional conformal black holes. In section \ref{Sec:TI}, we analysis and synthesize the curves of specific heat capacity. Finally, we summarize our results in section \ref{Sec:Con}.
	
	\section{Thermodynamic Properties of CG Holography in Four Dimensions}
	
	\label{Sec:CGH}
	
	In this section, the thermodynamic properties of four-dimensional conformal gravity holography are discussed briefly. Conformal gravity is a theory that remains unaltered under the Weyl transformation, which can be deduced from the action of the Weyl conformal curvature \cite{Hall:2008zzc}. As the Weyl action accompanied with the electromagnetic field, one can obtain \cite{Riegert:1984zz}
	\begin{equation}\label{eqn:Q1}
		\mathcal{I}=-\frac{1}{4}\int\sqrt{-g}d^{4}x(\beta C_{abcd}C^{abcd}+F^{cd}F_{cd}), \\
	\end{equation}
	where $C_{abcd}$ refers to conformal curvature tensor and $F_{ab}\equiv A_{a,b}-A_{b,a}$ refers to electromagnetic field tensor. By changing $g_{ab}$ and $A_{b}$, the most static as well as spherically symmetric solution can be given by
	\begin{equation}\label{eqn:Q2}
		ds^{2}=-\mathcal{F}(r)dt^{2}+\frac{dr^{2}}{\mathcal{F}(r)}+r^{2}(d\theta^{2}+sin^{2}\theta d\phi^{2}). \\
	\end{equation}
	
	The metric function $\mathcal{F}(r)$ has its general form written as follow
	\begin{equation}\label{eqn:Q3}
		\mathcal{F}(r)=\lambda r^{2}+\mu r+\delta+\frac{\eta}{r}, \\
	\end{equation}
	where the constants $\lambda$, $\mu$, $\delta$, $\eta$, are discovered to satisfy
	\begin{equation}\label{eqn:Q4}
		3\mu\eta-\delta^{2}+1+\frac{3q^{2}}{2\beta}=0. \\
	\end{equation}
	
	In 2010, Grumiler developed an effective model for gravity using in the large-scale central object and found a solution consisting of the cosmological constant as well as the Rindler parameter $a$, where $a$ is used for setting the physical scale and the subdominant terms. Subsequently, in 2014, he proposed a solution for a form of conformal gravity without taking Maxwell's field into account \cite{Grumiller:2010bz}
	\begin{equation}\label{eqn:Q5}
		\mathcal{F}(r)=\sqrt{1-12aM}-\frac{2M}{r}+2ar-\frac{\Lambda}{3}r^{2}, \\
	\end{equation}
	where $a$, $\Lambda$, $M$ are devoted as the Rindler parameter, the cosmological constant and the mass of black hole respectively. In the extended phase space, the cosmological constant and the thermodynamic pressure are satisfied with the following relation
	\begin{equation}\label{eqn:Q6}
		P=-\frac{\Lambda}{8\pi}.
	\end{equation}
	
	Therefore, the metric function is rewritten as
	\begin{equation}\label{eqn:Q7}
		\mathcal{F}(r)=\sqrt{1-12aM}-\frac{2M}{r}+2ar+\frac{8\pi P}{3}r^{2}.
	\end{equation}
	
	The radius of horizon ($r_{+}$) can be calculated from $\mathcal{F}(r_{+})=0$ while the temperature can be given by
	\begin{equation}\label{eqn:Q8}
		T=\frac{1}{4\pi r_{+}}[ar_{+}+8\pi Pr_{+}^{2}+\sqrt{1-3a^{2}r_{+}^{2}-16\pi aPr_{+}^{3}}].
	\end{equation}
	
	The mass of black hole is solved in the same way as the previous equation for the radius of horizon and then the solution is obtained by
	\begin{equation}\label{eqn:Q9}
		M=\frac{r_{+}}{2}[\sqrt{1-3a^{2}r_{+}^{2}-16\pi aPr_{+}^{3}}-ar_{+}+\frac{8\pi P}{3}r_{+}^{2}].
	\end{equation}
	
	Meanwhile, the thermodynamic volume calculated from $V=(\frac{\partial M}{\partial P})_{S}$ can be written as
	\begin{equation}\label{eqn:Q10}
		V=\frac{4}{3}\pi r_{+}^{3}[1-\frac{3ar_{+}}{\sqrt{1-3a^{2}r_{+}^{2}-16\pi aPr_{+}^{3}}}].
	\end{equation}
	
	Its entropy is able to be computed by employing
	\begin{equation}\label{eqn:Q11}
		S=\frac{\mathcal{A}}{4l^{2}}=\frac{8\pi^{2}}{3}Pr^{2},
	\end{equation}
	where $\mathcal{A}$ and $l$ is defined as $\mathcal{A}=4\pi r^{2}$ and $l=\sqrt{\frac{3}{8\pi P}}$ respectively.
	
	Eq. {$\left(\ref{eqn:Q8}\right)$}, Eq. {$\left(\ref{eqn:Q9}\right)$}, Eq. {$\left(\ref{eqn:Q10}\right)$} and Eq. {$\left(\ref{eqn:Q11}\right)$} satisfy the first law of thermodynamics, which is denoted as
	\begin{equation}\label{eqn:Q12}
		dM=TdS+VdP+\chi da,
	\end{equation}
	where $\chi$ is a physical quantity with respect to the parameter $a$.
	
	Therefore, via the definition of total derivative, $\chi$ can be calculated by
	\begin{equation}\label{eqn:Q13}
		\chi=(\frac{\partial M}{\partial a})_{S,P}=-\frac{r_{+}}{2}[1+\frac{3r_{+}(a+\frac{8\pi P}{3}r_{+})}{\sqrt{1-3a^{2}r_{+}^{2}-16\pi aPr_{+}^{3}}}].
	\end{equation}
	
	It represents the modification in the first law of thermodynamics for extended phase space caused by the Rendler parameter $a$.

	As investigating four-dimensional conformal black holes, we have also analysed its reverse isoperimetric inequality. The ratio $\mathcal{R}$ is given by
	\begin{equation}\label{eqn:Q14}
		\mathcal{R}=(\frac{3V}{4\pi})^{\frac{1}{3}}(\frac{4\pi}{\mathcal{A}})^{\frac{1}{2}}.
	\end{equation}
	
	Thus, we can obtain the ratio $\mathcal{R}$ of four-dimensional conformal black hole
	\begin{equation}\label{eqn:Q15}
		\mathcal{R}=[1-\frac{3ar_{+}}{\sqrt{1-3a^{2}r_{+}^{2}-16\pi aPr_{+}^{3}}}]^{\frac{1}{3}}.
	\end{equation}
	
	Despite general black holes satisfy the inverse isoperimetric inequality ($\mathcal{R}\geqslant1$), there is still a special class that satisfy the case of $\mathcal{R}\text{<}1$, which is called super-entropy black hole. Combining with Eq. {$\left(\ref{eqn:Q15}\right)$}, it can be clearly found that a four-dimensional conformal black hole violates the inverse isoperimetric inequality which indicates that it’s a type of super-entropy black hole. Hence, combined with the conjecture of \cite{Cong:2019bud} , we further investigate the specific heat capacities of the four-dimensional conformal black holes.

	\section{Thermodynamic Instabilities}
	
	\label{Sec:TI}
	In this section, the specific heat capacities of the four-dimensional conformal black hole are analysed. Firstly we calculate the specific heat capacity at constant pressure $C_{P}$ exploiting the following equation
	\begin{equation}\label{eqn:Q18}
		C_{P}=T\frac{\partial S}{\partial T}\mid_{P}.
	\end{equation}
	
	The analytical solution of $C_{P}$ can be obtained via combining  Eq. {$\left(\ref{eqn:Q8}\right)$} as well as  Eq. {$\left(\ref{eqn:Q11}\right)$}. As for the specific heat capacity at constant volume $C_{V}$, its specify analytical solution can be written by using $C_{P}$ \cite{Johnson:2019mdp}, which is calculated as follow
	\begin{equation}\label{eqn:Q19}
		C_{V}=C_{P}-TV\alpha_{P}^{2}\kappa_{T},
	\end{equation}
	where $\alpha_{P}\equiv V^{-1}\frac{\partial V}{\partial T}\mid_{P}$ is the isobaric expansion coefficient and $\kappa_{T}\equiv-V\frac{\partial P}{\partial V}\mid_{T}$ is the isothermal bulk modulus.
	
	We notice that the temperature need to keep constant during the calculation of $\kappa_{T}$. However, in practical calculations, it is almost impossible to introduce temperature into the volume directly. Hence we take advantage of the fact that the temperature and mass of four-dimensional conformal black hole are composed of the same variables, which enlightens us that sustaining the temperature constant means remaining the mass constant. Thus, we use $\kappa_{T}\equiv-V\frac{\partial P}{\partial V}\mid_{M}$ to obtain $\kappa_{T}$. Eventually via this method, an explicit solution for $C_{V}$ is obtained. However, the formula is extremely long so that it is not represented in the paper.
	
\begin{figure}
		\begin{center}
			\subfigure[{}]{
				\includegraphics[width=0.48\textwidth]{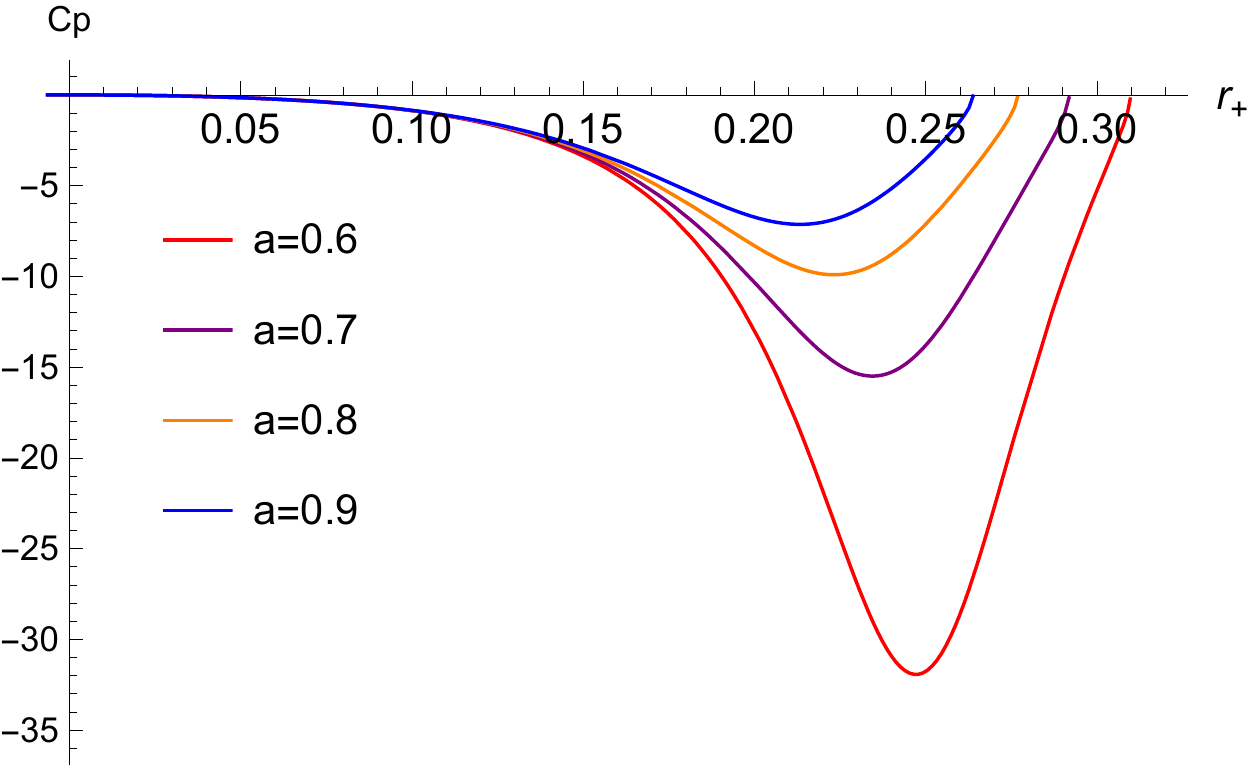}\label{fig:Cpa69}}
			\subfigure[{}]{
				\includegraphics[width=0.48\textwidth]{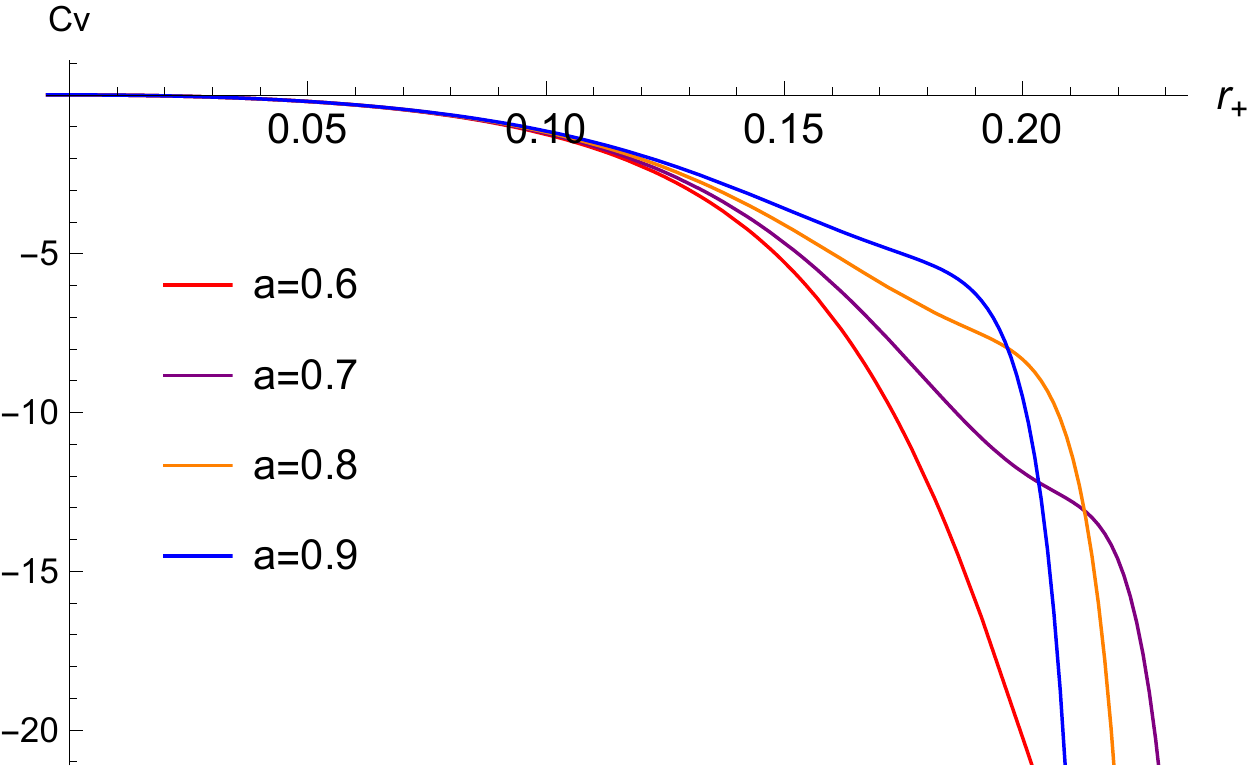}\label{fig:Cva69}}
		\end{center}
		\caption{The curves of specific heat capacity plotted with $a=0.6$, $0.7$, $0.8$and $0.9$, where $P=1$  }%
		\label{fig:C01}
	\end{figure}

	In Fig.\ref{fig:C01}, the $C_{P}$ curves plotted with different values of $a$ are presented in the left. Whereas, the $C_{V}$ curves are shown in the right. When the value of $a$ ranges from $0.5$ to $1$, it can be obviously observed that the $C_{V}$ curves are always negative, so do the $C_{P}$ curves. In thermodynamics, $C_{V}<0$ denotes that the system is unstable. Thus with the parameters chosen in Fig.\ref{fig:C01}, the four-dimensional conformal black hole is thermodynamically unstable, which obeys the conjecture in \cite{Johnson:2019mdp}.
	
\begin{figure}
		\begin{center}
			\subfigure[{}]{
				\includegraphics[width=0.48\textwidth]{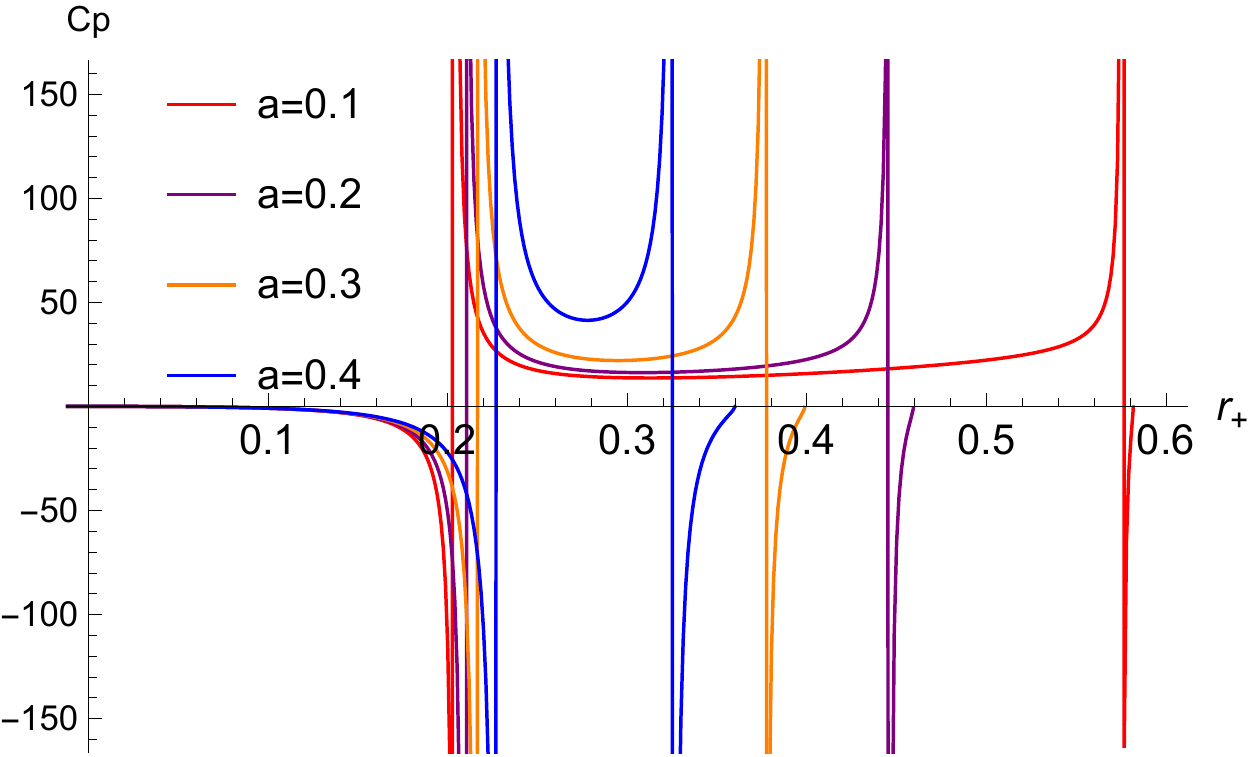}\label{fig:Cpa14}}
			\subfigure[{}]{
				\includegraphics[width=0.48\textwidth]{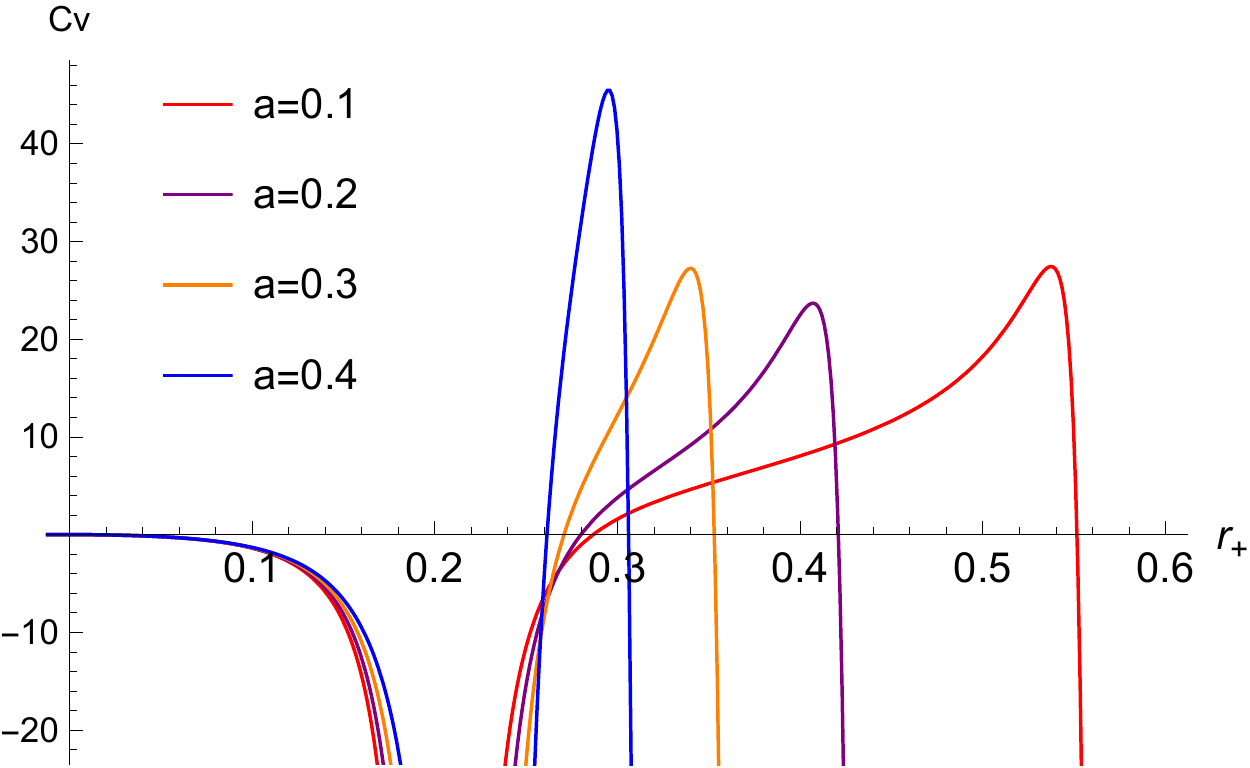}\label{fig:Cva14}}
		\end{center}
		\caption{The curves of specific heat capacity plotted with $a=0.1$, $0.2$, $0.3$and $0.4$, where $P=1$}%
		\label{fig:C02}
	\end{figure}

	In Fig.\ref{fig:C02}, the $C_{P}$ curves plotted with different values of $a$ are presented in the left. Whereas, the $C_{V}$ curves are shown in the right. The only difference of calculation between Fig.\ref{fig:C01} and Fig.\ref{fig:C02} is the changed range of values for $a$. In Fig.\ref{fig:C02}, the value of $a$ have been restricted to the range from $0$ to $0.5$.
	
\begin{figure}
		\begin{center}
			\subfigure[{}]{
				\includegraphics[width=0.48\textwidth]{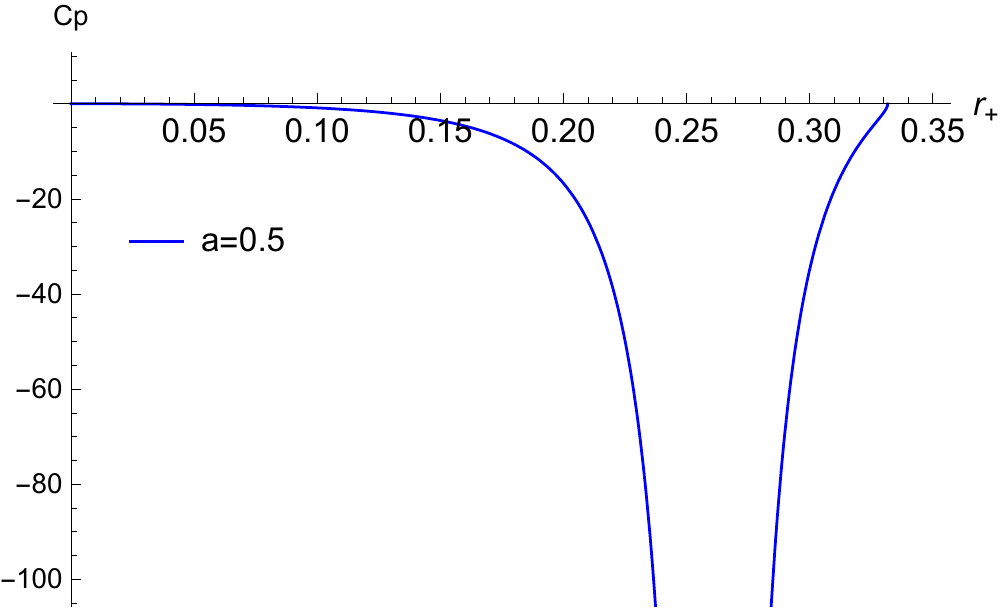}\label{fig:Cpa5}}
			\subfigure[{}]{
				\includegraphics[width=0.48\textwidth]{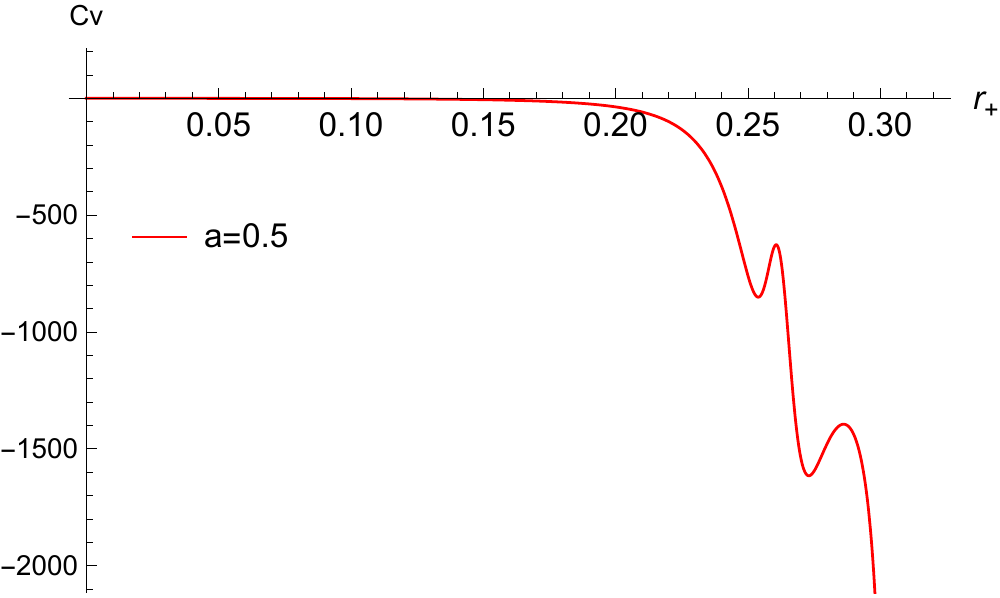}\label{fig:Cva5}}
		\end{center}
		\caption{The curves of specific heat capacity plotted with $a=0.5$, where $P=1$}%
		\label{fig:C03}
	\end{figure}

	With the comparison between Fig.\ref{fig:C01} and Fig.\ref{fig:C02}, we find differences in the curves using $a=0.5$ as a boundary. Thus observing the curve of $C_{P}$ as well as $C_{V}$ for $a=0.5$ plotted in Fig.\ref{fig:C03}, we can clearly find that the case of $a=0.5$ differs from both $0<a<0.5$ and $0.5<a<1$, which resembles a state in-between the two.

	Comparing the Fig.\ref{fig:C01} and Fig.\ref{fig:C02}, it is straightforward to see how the trend as well as shape of the curves change. And there is an interesting part coming in. We find a non-negative part of the $C_{V}$ curves when the value of $a$ is in the range from $0$ to $0.5$, which defies the conjecture in \cite{Johnson:2019mdp}. In combination with the conjecture proposed in \cite{Cong:2019bud}, we then plot the curves of $C_{P}$ in a four-dimensional conformal black hole. In contrast to what is described in \cite{Cong:2019bud}, the region that we observe as non-negative according to the $C_{V}$ curves does not exactly correspond to $C_{P}<0$ in the $C_{P}$ curves.

	With these investigations, we end up with a counterexample that contradict the conjectures.

	\section{Conclusion}
	
	\label{Sec:Con}
	
	In this section, a summary of the paper is presented. We derived and calculated the specific heat capacities of the four-dimensional conformal black hole and on the basis of which the corresponding curves of $C_{P}$ as well as $C_{V}$ were plotted. We have shown that both conjectures in \cite{Cong:2019bud} and \cite{Johnson:2019mdp} are violated in the range of $a$ from 0 to 0.5 for the four-dimensional conformal black holes. Not only do four-dimensional conformal black holes have positive values of $C_{V}$, but also the same region of radius exists the positive values of $C_{P}$. It is impossible to explain whether the super-entropy is linked to thermodynamic instabilities in that particular region. Thus far, we have found a counterexample that defies the conjectures in \cite{Cong:2019bud,Johnson:2019mdp}.

	\section{\noindent\textbf{Acknowledgements }}
	We are grateful to Yiqian He, Deyou Chen, Rui Yin, Jing Liang, Peng Wang, Haitang Yang, Jun Tao and Xiaobo Guo for useful discussions. This work is supported in part by NSFC (Grant No. 11747171), Xinglin Scholars Project of Chengdu University of Traditional Chinese Medicine (Grant no.QNXZ2018050).

\end{document}